\documentclass[final]{IEEEtran}
\usepackage{cite}
\usepackage{graphicx}
\usepackage{picinpar}
\usepackage[cmex10]{amsmath}
\usepackage{amsmath,amsfonts,amssymb}
\usepackage{subfigure}
\usepackage{amsthm}
\usepackage{algorithm}
\usepackage{algorithmic}
\usepackage{stfloats}
\usepackage{bm}
\usepackage{color}
\usepackage{booktabs}
\usepackage{makecell}
\usepackage{mathrsfs}
\usepackage{stfloats}
\usepackage{url}
\usepackage{multirow}
\usepackage{threeparttable}
\usepackage{lipsum}
\usepackage{epstopdf}
\DeclareMathAlphabet{\mathbbold}{U}{bbold}{m}{n}
\begin{document}
\pdfoptionpdfminorversion=6
\newtheorem{lemma}{Lemma}
\newtheorem{corol}{Corollary}
\newtheorem{theorem}{Theorem}
\newtheorem{proposition}{Proposition}
\newtheorem{definition}{Definition}
\newcommand{\e}{\begin{equation}}
\newcommand{\ee}{\end{equation}}
\newcommand{\eqn}{\begin{eqnarray}}
\newcommand{\eeqn}{\end{eqnarray}}
\renewcommand{\algorithmicrequire}{ \textbf{Input:}}
\renewcommand{\algorithmicensure}{ \textbf{Output:}}
\title{LEO Satellite Constellations for 5G and Beyond: How Will They Reshape Vertical Domains?}

\author{Shicong Liu, Zhen Gao\textsuperscript{*}, Yongpeng Wu, Derrick Wing Kwan Ng, Xiqi Gao, Kai-Kit Wong, Symeon Chatzinotas, and Bj\"{o}rn Ottersten
\thanks{\textsuperscript{*}: Corresponding author.}
}

\maketitle

\begin{abstract}
The rapid development of communication technologies in the past decades has provided immense vertical opportunities for individuals and enterprises. However, conventional terrestrial cellular networks have unfortunately neglected the huge geographical digital divide, since high bandwidth wireless coverage is concentrated to urban areas. To meet the goal of ``connecting the unconnected'', integrating low Earth orbit (LEO) satellites with the terrestrial cellular networks has been widely considered as a promising solution. In this article, we first introduce the development roadmap of LEO satellite constellations (SatCons), including early attempts in LEO satellites with the emerging LEO constellations. Further, we discuss the unique opportunities of employing LEO SatCons for the delivery of integrating 5G networks. Specifically, we present their key performance indicators, which offer important guidelines for the design of associated enabling techniques, and then discuss the potential impact of integrating LEO SatCons with typical 5G use cases, where we engrave our vision of various vertical domains reshaped by LEO SatCons. Technical challenges are finally provided to specify future research directions.

\end{abstract}

\section{Introduction}\label{S1}
\IEEEPARstart{T}{he} unprecedented development of communication technology in the past few decades has demanded an exponential growth in the overall transmission rate\cite{5G,6G}.
Unlike previous mobile communications, from 1G to 4G, 5G networks aim for ubiquitous connections to help digitize the economy and contribute towards the global digital transformation, rather than barely improving the overall data rate.
The driving force for further beyond 5G (B5G) even 6G research is the demand-intensive vertical industries, which can be classified as one or a mix of the following usage scenarios: enhanced mobile broadband (eMBB), massive machine type communications (mMTC), and ultra reliable low latency communications (URLLC)\cite{5Gusage}.

Affected by the COVID-19 pandemic, vertical applications such as media, public safety, and eHealth have attracted widespread attention, and are expected to become a new normality in the future. According to the report of international telecommunication union (ITU), nearly half of the population still remains unconnected\cite{ITU}, whereas eliminating the blind areas by deploying much more terrestrial base stations (BSs) will impose a heavy burden on the telecom operators. On the other hand, the coverage of non-deployable areas such as aero and maritime lanes remains to be solved. Against the aforementioned constraints, space-based satellite constellations (SatCons) emerge as the times require.

With the emerging concept referred to as mega-constellations\cite{mega_con}, the era of seamless connectivity is more realistic than ever. Tremendous amount of satellites in low Earth orbit (LEO), medium Earth orbit (MEO), and geostationary Earth orbit (GEO) can collaborate together forming SatCons, which completely changes the static topology in conventional terrestrial networks and enables flexible network deployment. Despite the fact that $3$ GEO satellites are enough to provide seamless coverage across the globe, the associated latency and low area spectral efficiency are generally intolerable in most 5G application scenarios. Hence, exploiting LEO satellites becomes a better option as operating on a lower orbit provides lower latency and higher service density.

In recent years, new launching and propulsion technologies have drastically reduced the cost, and advanced volume manufacturing technologies have paved the way for the deployment of mega constellations, especially for LEO satellites. For LEO, the lower orbital altitude leads to a lower latency for delay-sensitive tasks such as video calls and streaming, while the high density constellation guarantees service capacity globally. Therefore, the amalgamation of satellite access and conventional terrestrial networks is an inevitable convergence for B5G and even 6G in meeting the increasing demands.

After this introduction, the article first introduces the roadmap of LEO SatCons development, after which the opportunities of employing LEO SatCons are discussed. We further discuss the technical challenges and highlight the research directions, and finally summarize our conclusion.

\section{Development Roadmap of LEO SatCons}
The idea of communicating via commercial LEO satellites dates back to the early 1990s. Since the earliest launch in 1957, satellites used for data transmission are mainly GEO satellites and the majority of them served as data broadcasting satellites, whose applications include satellite TV programs, positioning, etc.
With the increasing demands for higher system capacity and lower latency, the concept of constellation composed of LEO satellites has been proposed\cite{mega_con}. Early designs of LEO satellites typically included a small number of satellites to ensure that every part of Earth is covered by at least one satellite.
In this section, we provide a comprehensive overview and comparison on the existing SatCons. Table \ref{comparison} gives a brief summary of the most well-known commercial satellite communication systems that utilize constellations to support wide-area coverage.
\renewcommand\arraystretch{1.25}
\begin{table*}[t]
\centering
\caption{Comparison of selected constellations}
\label{comparison}
\begin{threeparttable}
\begin{tabular}{c|c|c|c|c|c|c}
\hline
{\bf Constellation}  & {\bf Regime} & {\bf Orbital height }& {\bf  Quantity }&{\bf  Bands } & {\bf Services}  &{\bf  Est. data rate}\\
\hline
Iridium Gen. 1  & LEO & $781$ km & 66 & L & Voice, data & 2.4 Kbps  \\
\hline
Globalstar  & LEO & $1414$ km& 48 & S, L  & Voice, data & $\sim$9.6 Kbps \\
\hline
Orbcomm Gen. 1  & LEO & $700\sim 800$ km & 36 & VHF & IoT \& M2M$^*$ communication & 2.4 Kbps\\
\hline
Skybridge  & LEO & $1457$ km & 64 & Ku  & Broadband Internet& 60 Mbps\\
\hline
Teledesic  & LEO & $1375$ km & 288(840) & Ka  & Broadband Internet& 64 Mbps \\
\hline\hline
Iridium NEXT  & LEO & $781$ km & 66 & L, Ka & Voice, data & 1.5 Mbps, 8 Mbps  \\
\hline
Orbcomm Gen. 2  & LEO & $700\sim 800$ km & 18 & VHF& IoT \& M2M communication & 4.8 Kbps\\
\hline
O3b  & MEO & $8063$ km & 20 & Ka& Broadband Internet & 500 Mbps\\
\hline
OneWeb  & LEO & $1200$ km & 648 & Ku & Broadband Internet & 400 Mbps\\
\hline
Starlink Gen. 1 & \multirow{2}{*}{LEO} & $335.9\sim 570$ km & 11926 & V & \multirow{2}{*}{Broadband Internet} & \multirow{2}{*}{100 Mbps}\\
Starlink Gen. 2 &  & $328\sim 614$ km & 30000 &Ku, Ka, E &  & \\
\hline
Telesat Phase. 1  & \multirow{2}{*}{LEO} & \multirow{2}{*}{$1015\sim 1325$ km} & 298 & \multirow{2}{*}{Ku, Ka} & \multirow{2}{*}{Broadband Internet} & \multirow{2}{*}{-}\\
Telesat Phase. 2  &   &   & 1373 & & & \\
\hline
Hongyan  & LEO & $1100$ km & 320 & L, Ka & Voice, broadband Internet & 100 Mbps\\
\hline
Kuiper  & LEO & $590\sim 630$ km & 3236 & Ka & Broadband Internet & -\\
\hline
\end{tabular}
\begin{tablenotes}
        \footnotesize
        \item[*$\ \ $]: Internet of Things and machine to machine.
      \end{tablenotes}
\end{threeparttable}

\end{table*}

\subsection{Overview of Early LEO Satellites}

In the late 20$^{\rm th}$ century, 
network operators and companies have made their initial attempts to deploy LEO constellations. Although most of the early attempts ended with bankruptcy, their experiences provide valuable lessons for the ventures.

{\bf{Iridium}} went into operation by Motorola in 1998 as one of the earliest LEO constellations adopting the Walker constellation pattern. 11 satellites are uniformly distributed on 6 orbital planes with a height of $780$ km, and are able to work in both time division multiple access (TDMA) and frequency division multiple access (FDMA) modes.

Different from Iridium, {\bf{Globalstar}} adopted the Walker Delta pattern, which is also known as Rosette, achieved the best coverage diversity at mid-latitudes, but no coverage at poles, and became commercially available in late 1998. 48 LEO satellites are designed at the altitude of $1414$ km adopting code division multiple access (CDMA) for satellite phones and narrow band data communication services.

{{\bf Teledesic}} also started its operation in 1998 and initially attempted to build the largest constellation with $840$ satellites at the altitude of $700$ km supporting the space-based broadband Internet access. After several reorganizations and merging, Teledesic progressively scaled down the number of its satellites to no more than $288$.

The past verticals were underdeveloped to nurture a mature LEO commercial operation in the last century. Therefore, in the beginning of $21$st century, the aforementioned early trials eventually lost their business in competition with the relatively mature terrestrial networks.

\subsection{State-of-the-Art Emerging Constellations}

Current flourishing verticals are booming a huge LEO market, which can further fuel the growth of the verticals. Particularly, with the development of terrestrial networks gradually approaching the bottleneck period, the unresolved vertical issues are left to the emerging SatCons.

{\bf Starlink} intends to launch around $42000$ LEO satellites, which make up the largest constellation ever, on several different altitudes of orbits. At the initial stage of arrangement, around 12000 satellites with 2 groups of orbital height ($335.9\sim 345.6$ km for $7518$ satellites and $540\sim 570$ km for $4408$ satellites) will be deployed, whereas 30000 more satellites awaiting approval will be deployed at the final stage. 
As an emerging SatCon, it is the first constellation that has applied for the use of V-band and E-band to support its advanced services in broadband Internet access globally, including commercial, residential, and governmental scenarios.

\begin{figure}[t]
	\centering
	\includegraphics[scale=0.35]
	{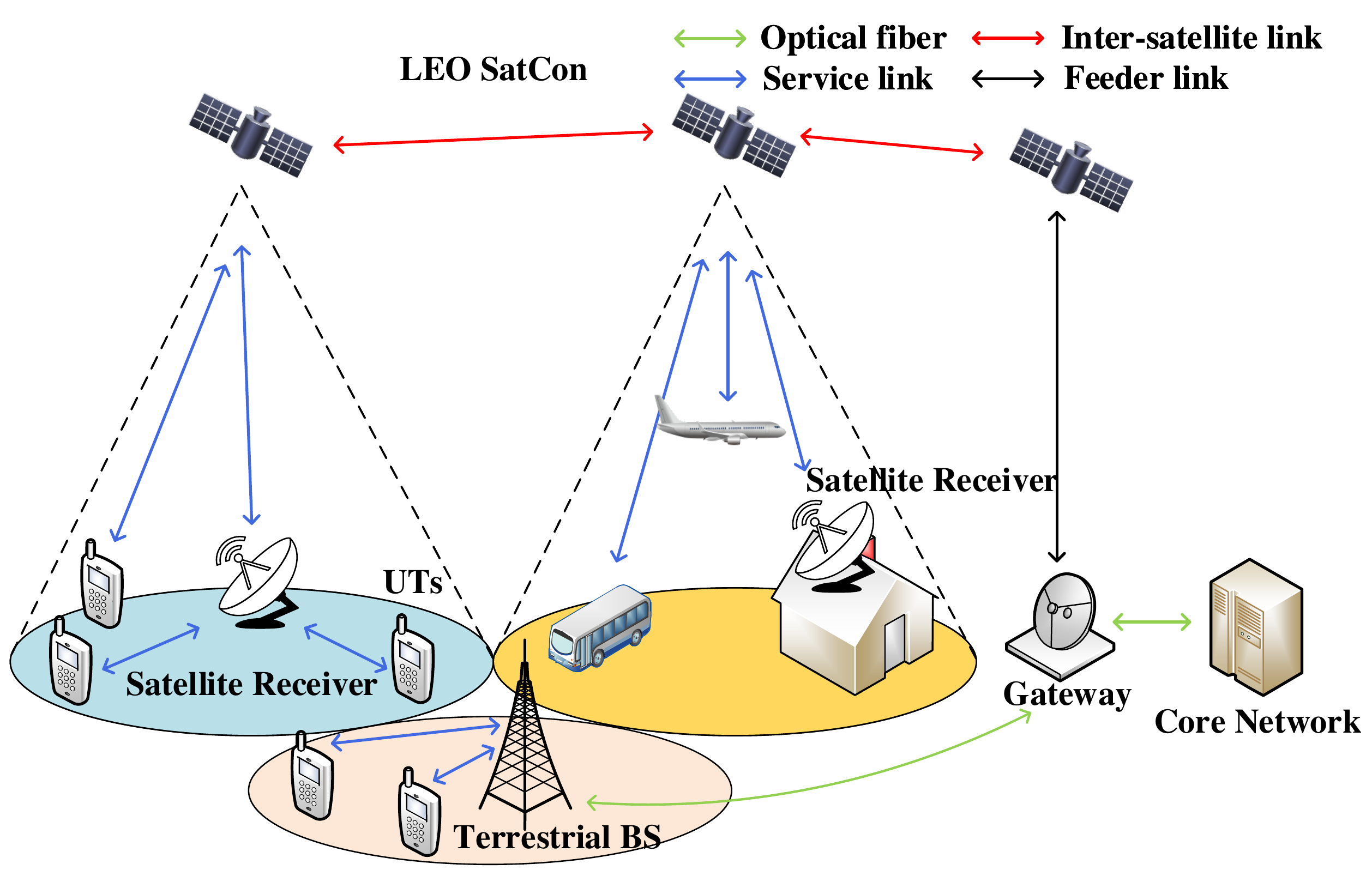}
	\caption{A schematic diagram of the network architecture of LEO SatCons and the common UT-SatCon access modes.}\label{accessMode}
\end{figure}

Apart from Starlink, several developing or designing LEO SatCons are in progress. {\bf OneWeb}$\footnote{OneWeb also went bankrupt in March 2020 due to the high cost of manufaturing and launching satellites, while seeking to expand the constellation to up to $48000$ satellites after being taken over by the new owners. }$ targets at seamless global coverage with customer demonstrations and commercial services in the upcoming future, and has launched 74 of 648 planned satellites to date. It is expected to provide up to $400$ Mbps downlink and $30$ Mbps uplink rate through special user terminals (UTs) or gateways. 
On the other hand, {\bf Telesat} also seeks to provide global coverage and high throughput broadband capacity on LEO constellation in polar and inclined orbits. Its first LEO satellite was launched in 2018, and is now supporting live demonstrations of satellite services across a variety of markets and applications.
The project {\bf Kuiper} is an initiative that proposes to build an LEO constellation with $3236$ satellites for providing broadband connectivity to the uncovered users. The related detail of other constellation competitors can also be found in Table \ref{comparison}. Moreover, we present the typical network architecture of LEO SatCons and common UT-SatCon access modes in Fig. \ref{accessMode}. 

In recent years, integrating 5G with LEO SatCons has become a trending topic. The low altitude characteristic of LEO SatCons reduces latency, and the massive deployment achieves global coverage with high service density, which can be an innovative component in building future wireless networks.

\section{Opportunities of LEO SatCons for 5G and Beyond}

In this section, we will provide the motivation of adopting LEO constellations, and discuss the impact on typical use cases as well as how they will reshape the vertical domains.
\subsection{Motivation of Integrating LEO SatCons}

To explore the unique advantages of applying LEO SatCons over terrestrial 5G networks, we first discuss the key performance indicators (KPIs) of space-based wireless network. 
These unique advantages motivate us to integrate LEO SatCons with B5G systems to achieve better performances. Specifically, these KPIs are discussed as follows.
\subsubsection{Latency} The orbital altitude determines the end-to-end communication latency. 
Extremely high altitude brings wide coverage, while the round-trip propagation latency (e.g., up to $240$ ms via GEO satellites) becomes intolerable under the requirements of current 5G standard. By contrast, the altitude of some of the latest constellations such as Starlink is as low as $320$ km. As a result, the corresponding round-trip air interface latency is reduced to around $3$ ms, which well meets the control plane round-trip latency requirements for most eMBB and mMTC usage, especially the IoT-based vertical scenarios.
\subsubsection{Reliability} Reliability is usually measured by the error rate of communication links, and is crucial to various vertical applications (e.g., eHealth and finance).
In addition to error rate, outage time is another manifestation of reliability in eMBB and URLLC scenarios. As the scale of constellations increases, users can be served by more satellites simultaneously, which significantly improve the reliability thanks to the spatial diversity gains.
\subsubsection{Data rate} In current standards, data rate for digital video broadcasting is around $30$ Mbps, while the customized streaming rate is far limited, which can be seen in early attempts of satellite voice services shown in Table. \ref{comparison}. In recent years, the data rate of customized streaming supported by constellations is increasingly required, hence it is important to reconsider the position of data rate in SatCon constructions.

\subsubsection{Availability} Availability is one of the most important KPIs for satellite communications, which refers to the probability that at least one satellite is visible to the UT. Deploying more LEO satellites implies an improved availability, which will significantly contribute to the quality of service (QoS).

\subsubsection{Cost} Cost is arguably the biggest challenge to profitability and long-term viability. Due to the financial burden from manufacturing, launching, operating, and so on, several companies have gone bankrupt in the early 21$^{\rm st}$ century. Fortunately, the recently developed satellite technologies and the soaring demand for ubiquitous broadband Internet access with promising market profits have fueled the development of affordable and healthy operation of LEO constellations.

\subsubsection{Mobility}
Due to higher frequency bands and harsh terrain, smaller cellular coverage and frequent migration via public transportation have caused mobility issues. The problem of mobility management in SatCons is more complicated than that of terrestrial networks, since the mobile entities include not only the UTs but also the high-speed moving satellites, which will pose the paging and handover issues as well as dynamic varying Doppler shifts. Therefore, effective mobility management is another important KPI for LEO SatCons.

\subsubsection{Connectivity density} An unprecedented number of satellites can be exploited to support high density of services, especially in rural areas. In recent year, the massive number of machine-type nodes has been increasing rapidly, which makes the connection density as a KPI. However, the terrestrial BSs are insufficient in rural areas and pathless areas, where the low-cost satellite receivers can be deployed to accommodate massive connectivity.

\subsubsection{Survivability}Fragile optical cables in terrestrial cellular networks can be easily destroyed by natural disasters, leading to frequent network paralysis. By contrast, the SatCons have the superiority of survivability against natural disasters owing to the highly flexible topology and wireless access, which plays a vital role in disaster recovery.

\subsection{Impact on B5G Use Cases and Future Visions}

\begin{figure*}[t]
     \centering
     \includegraphics[width=15cm]
     {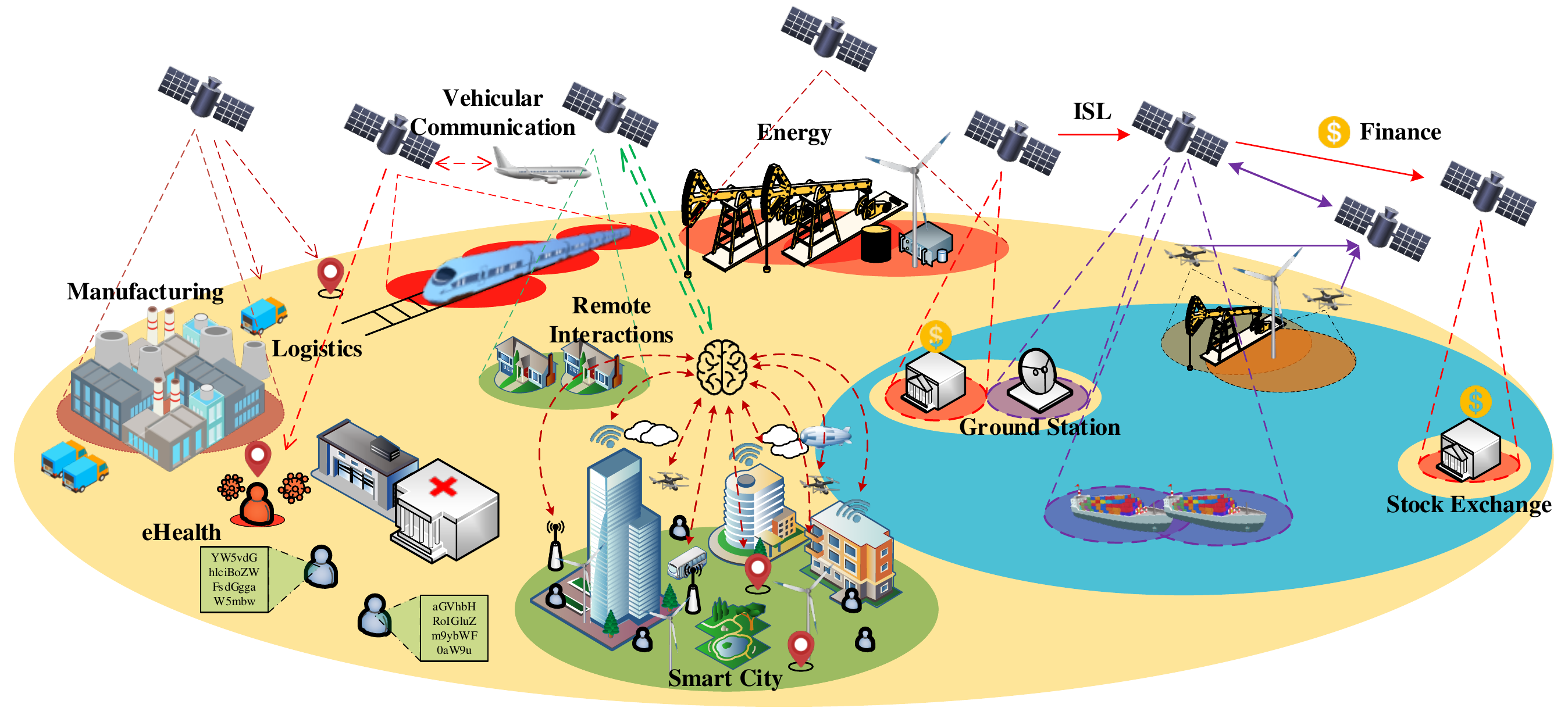}
     \caption{Ambitious vision of various vertical domains reshaped by LEO SatCons.}\label{satelliteCommSys}
\end{figure*}

As mentioned before, the huge geographical digital divide issue\cite{divide} caused by harsh terrain environments in conventional terrestrial cellular networks needs to be addressed. B5G or even 6G integrated with the space-based LEO constellations is envisioned to reduce the digital divide. Benefited from the lower latency, higher reliability and data rate, as well as denser connectivity and vast coverage, LEO SatCons are anticipated to reshape the traditional vertical domains and promise a myriad of opportunities for rising industries. Below we discuss the potential impact of LEO SatCons on typical use cases as shown in Fig. \ref{satelliteCommSys}.

\subsubsection{Industrial IoT}

Industrial IoT (IIoT), also known as Industry 4.0, is the combination of massive IoT nodes with industry applications. Basically, IIoT can be summarized as sensor-equipped industrial machines connected via private network with other supporting devices, and is suffering from piles of issues such as network topology, security, distributed manufacturing, personalized assembling, and cognitive supply networks. Overcoming these problems relies on fine-grained connectivity and high-accuracy positioning services, which have to require  massively deployed terrestrial BSs. By contrast, LEO SatCons is a viable approach to solve these challenges and an enabling technology for business-to-business (B2B) promotion.

\subsubsection{Agriculture}
Compared with industry, agriculture relies more on the convenience brought by the LEO SatCons. The cultivated land is typically located in the vast and sparsely populated areas, which brings the burden of cost for optical cables deployment. As more machine-type nodes are utilized in agricultural practice, deploying LEO satellites for crops and livestock monitoring and unmanned operations is an inevitable trend, and will significantly contribute to the productivity in agriculture.
\subsubsection{eHealth}
Public heath emergencies have stimulated global attention on medical services, which along with digitalization, virtualization, and decentralisation constitute the major trend towards eHealth. Among the most frequently cited issues being standardizing progressively in the roadmap, ubiquitous access and lifestyle evolution can be settled by LEO SatCons. With the intensification of decentralization, medical resources can be accessed from trusted entities such as SatCon operators. On the other hand, as the wearable devices become popularized, SatCons are still one of the most competitive solution for seamless health data monitoring and analysis in real time.

\subsubsection{Energy}
The newly discovered fossil energy is generally distributed in remote areas and maritime areas,  which are usually far from the 5G terrestrial cellular coverage. Nevertheless, LEO SatCon is the most suitable solution to support intelligent and remote operations. Most of the operation environment, such as the offshore plaform and the radioactive environment, is not compatible for long-term onsite in person operations. Therefore, the full information and necessary control via LEO constellations will greatly promote the development of energy fields. 
\subsubsection{Intelligent Vehicular Networks}
As the LEO SatCons become popularized thanks to the advanced manufacturing and low-cost launching techniques, elimination of coverage dead zones in air lines, sea lanes, and high-speed rails becomes a reality. Employing new antennas tailored for LEO SatCons, the airplanes, vessels, and high-speed trains can instantaneously become new dynamic nodes in the satellite network topology. Network coverage can be unprecedentedly wide, whereas the link performance such as reliability, capacity, and latency will be much better than before.
The enhanced vehicular network can fulfill the requirements raised by eMBB, and further makes it possible to deal with trading, streaming, and monitoring.
\subsubsection{Remote Interactions}
Remote interactions including education, online-work, online-entertainment and shopping for isolated and rural areas is not a novel idea, but it is not until the pandemic that these remote applications have gained most public attention. Global coverage via LEO SatCons with low latency and wide bandwidth paves the way for people in rural areas to interact remotely. Through allocating satellite resources, we can effectively promote the remote interactions as a safe and efficient solution in emergencies, a daily lifestyle, and even driving forces which conform to the unmanned fashion for labor-intensive industries.

\subsubsection{Finance}
Financial applications such as asset tracking (AT) and high-frequency trading (HFT) are typical scenarios for LEO satellites. AT requires one-way communication to achieve physical tracking, which is implemented by terrestrial networks and global positioning system (GPS). LEO satellites will improve the unreliability of terrestrial network and the positioning accuracy of GPS, therefore achieving better performances. HFT can be processed by computer algorithms rapidly in fractions of seconds and is sensitive to latencies. Space-based transmissions are expected to offer lower latency than undersea fiber, which may promote financial profit in the long term\cite{FSO}.

\subsubsection{Smart City}
The dominant issue for creating a smart city is the obsolete infrastructure, imperfect coordination mechanism, insufficient resources, and etc. In a typical smart city, massive IoT sensors distributed in cities can be directly connected to the ``City Brain'', where the cross-system multi-mode data are processed in real time. LEO satellites will play an important role in synchronous information exchange and dynamic resource allocation with fast response and high concurrency, and is envisioned to play a part in traffic management, large-scale activity flow monitoring, smart tourism, and cloud-edge integrated infrastructure management.

\section{Technical Challenges and Research Directions in LEO SatCons}

In this section, we introduce some key technologies that support various vertical applications mentioned above. The challenges are presented from the following three aspects.
\subsection{Network Architecture}
As shown in Fig. \ref{accessMode}, there are 2 types of methodologies that establish the constellation-based network architecture:
\begin{itemize}
\item UT directly communicates with LEO satellites, or through portable receivers.
\item UT communicates via terrestrial BSs, and the satellite-ground link serves as backhaul for data exchange.

\end{itemize}

The former strategy is adopted by Iridium, Globalstar, and some other operators by promoting their satellite UTs in the early stages, which has caused inconvenience and financial burden on both subscribers and operators. In contrast, the latter strategy still depends highly on the terrestrial stations to complete the network topology. 

In the past few years, the integration of spaceborne, airborne, and terrestrial networks has aroused widespread interest, where high altitude platforms and unmaned aerial vehicles can effectively enhance the SatCons' service coverage and QoS to accommodate various vertical scenarios. However, the network structure and routing protocols among different layers, as well as the heterogeneous network access problems for UTs among diversified devices still need to be further discussed.

\subsection{Mobility Management}
\begin{figure}[t]
	\centering
	\includegraphics[scale=0.45]
	{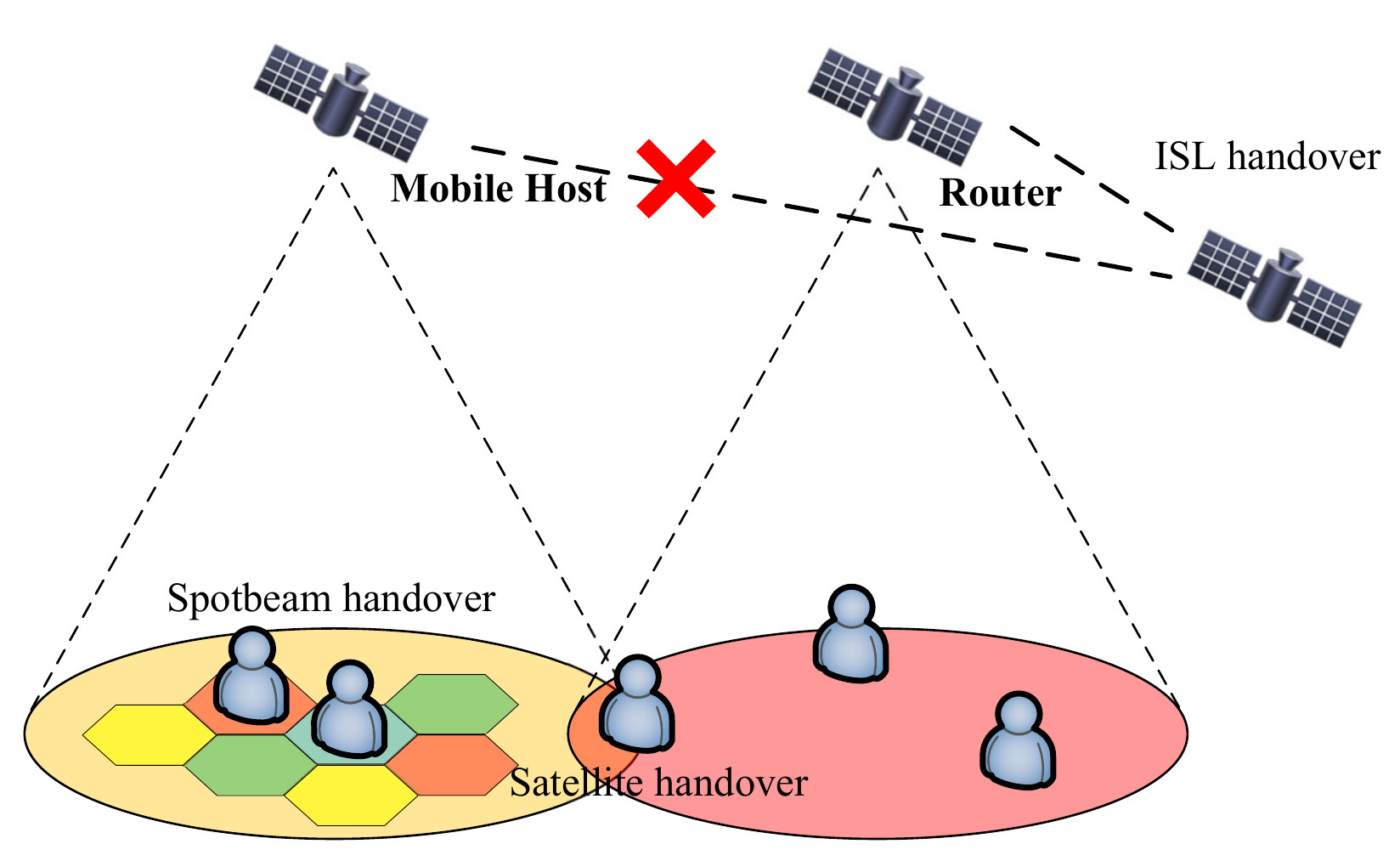}
	\caption{Typical link layer handover schemes.}\label{handover}
\end{figure}
Mobility management was first considered in cellular networks for supporting continuous communication applications.  Cells in conventional terrestrial cellular networks are statically tied to a radio access network (RAN). By contrast, spotbeams of LEO satellites are moving at high speed with a fixed trajectory. The cell has to be covered by different beams from different satellites in every time instant, since it is difficult for satellite based RAN to provide service continuously after the initial registrations. Besides, paging and handover procedures should be adapted to take into account the relative motion of the cell pattern with respect to the tracking area\cite{3GPP_NTN}. Particularly, handover techniques for link layer and network layer are urgently required for LEO satellite communication systems~\cite{handover, satellitehandover}.

\subsubsection{Link layer}
Link layer handover occurs when one or more links between the communication nodes have changed due to the dynamic connectivity patterns caused by high mobility of LEO satellites. This can be further classified into spotbeam handover, satellite handover, and inter-satellite link (ISL) handover as illustrated in Fig. \ref{handover}:

\begin{itemize}
\item
{\bf Spotbeam handover} happens when terrestrial nodes actively or passively cross the boundary between $2$ adjacent spotbeams of one satellite. 
\item
{\bf Satellite handover} happens when a node is no longer served by the previous satellite, and is transferred to another satellite.
\item
{\bf ISL handover} happens when an ISL is interrupted due to the change in distance or field-of-view angle, where new ISL establishes and causes rerouting issues.
\end{itemize}

\subsubsection{Network layer}
Network layer handover occurs when a satellite or terrestrial node needs to change its IP address. When satellites operate as mobile hosts that exchange data with terrestrial stations for different communication terminals, the IP address of the satellite is bonded with the terrestrial stations. When the satellite leaves the coverage area of the previous terrestrial station and starts to bond with another terrestrial station, the IP address of the satellite has to be updated, which requires network layer handover. Additionally, when the satellite operates as a router, users covered by these satellites may require a handover in the case of switching between spotbeams or satellite routers.

For handover and paging to be efficiently operated in LEO SatCons, future UT may be required to report its accurate location information periodically, and the ephemeris information of the LEO SatCons should be exploited to determine their footprints of each beam. Therefore, SatCons can select the best beams from each satellite to cover the users in demand. Moreover, the duration that UT would remain to be covered can also be estimated, which is convenient for SatCons to calculate the best candidate to handover. The adaptation of terrestrial protocols also needs further study to possibly utilize the location and ephemeris information in fixed trajectory LEO SatCons.

\subsection{Physical Layer Transmission}

Experiments for LEO SatCons are underway\cite{mega_con}, while the integration of satellite access with terrestrial 5G networks still remains unsettled. Commercial solutions for mobile satellite communication include setting up several terrestrial gateways, densely deploying tailored antennas, and using specified UTs. 
 
According to the current 5G technology and non-terrestrial network status, several incompatibilities still need to be resolved in future integrations.
\subsubsection{\bf Spectrum planning and interference management}
\begin{figure*}[t]
	\centering
	\includegraphics[width=11cm]
	{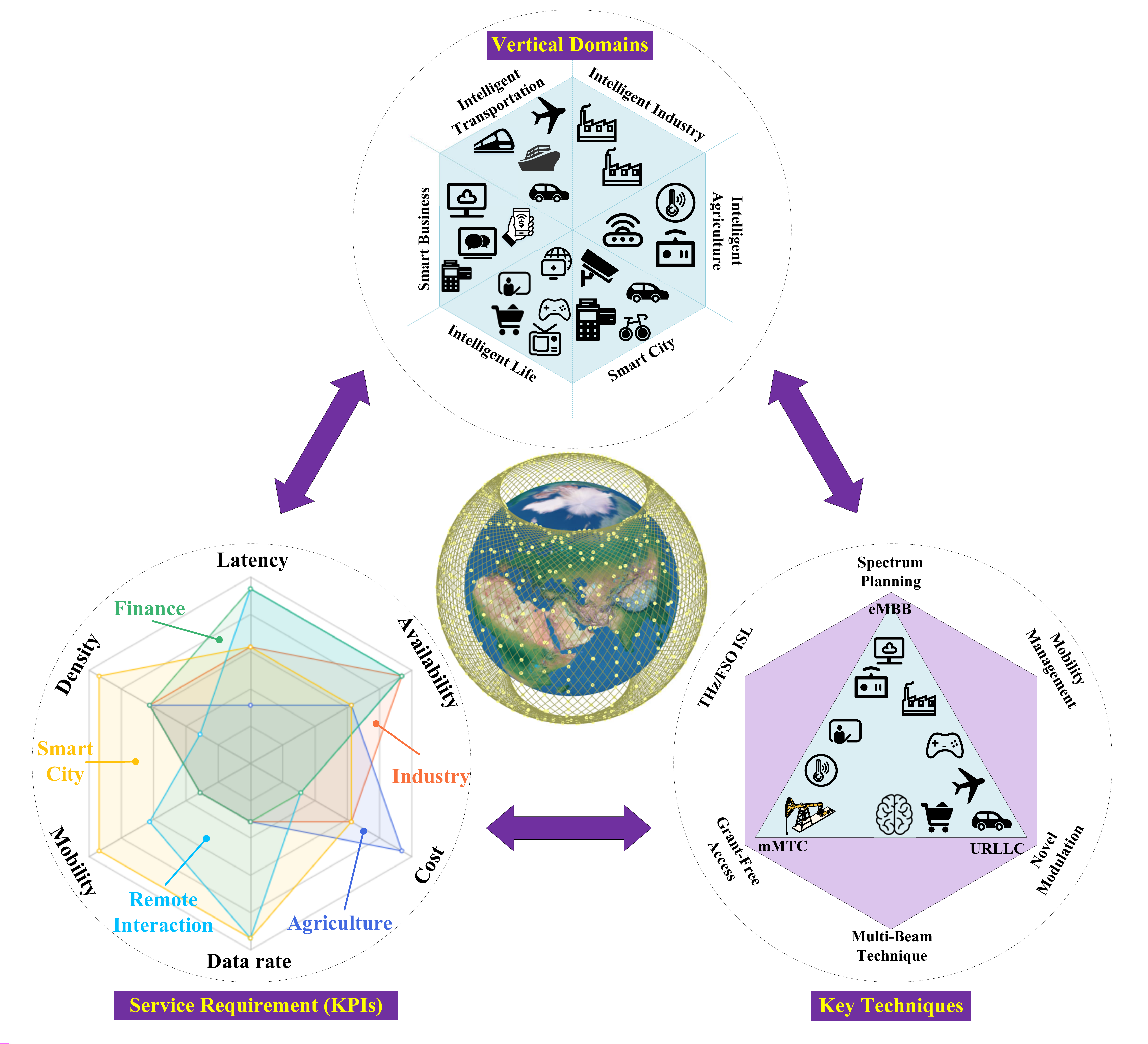}
	\caption{Relationship between vertical domains, service requirements, and key techniques.}\label{cycle}
\end{figure*}
5G has scheduled dozens of new frequency bands, which further squeeze the idle bands in sub-$6$G frequencies. The availability of tens of thousands of satellites in the near future will accelerate the exhaustion of limited spectrum, therefore pursuing higher frequency such as Ka bands has become a broad consensus as shown in Table \ref{comparison}. {Considering multiple verticals require large amount of spectrum resources, the interferences caused by frequency reuse and spectrum sharing among different orbital planes and terrestrial stations remain open issues, since the available radio spectrum is still costly for SatCon operators\cite{survey}. To aviod the undesired interference, flexible dynamic spectrum resource sharing schemes should be standardized to ensure robust interference management, and the advanced beamforming algorithms and digital signal processing techniques\cite{MIMO,NOMA} are demanded to handle interferences. 
} 

\subsubsection{\bf Advanced array and multi-beam technique}
{Massive multiple-input multiple-output (MIMO) has been applied in terrestrial cellular wireless networks, and recognized as one of the enabling technologies in 5G\cite{yljsac}. As the spaceborne stations are facing the increasing demand for availability, data rate, and reliability, massive MIMO with phased-array is expected to be exploited on SatCons to support flexible beamforming. On the other hand, active phased array and multi-beam antennas also require further research to be employed on UTs to make highly effcient use of spectrum resources and effectively track the beams from space-based stations with high mobility.
}

\subsubsection{\bf Terahertz and free space optical based ISL}
{The emerging data-intensive verticals promote the application of ISLs to accommodate unprecedented data traffic.}
Terahertz (THz) and free space optical (FSO) are two of the most promising techniques to support ISL due to their extremely high directivity, ultra-wide band, and security characteristics. In a turbulence-free outer space scenario, THz or FSO based ISLs can be  established for routing or data exchange without distortion, {therefore enhance the performance of SatCons. 
Note that optical frequencies can provide extremely high antenna gain at a relatively small antenna size, thereby effectively reduce the size and power consumption\cite{FSO}. Therefore, trending is that FSO based ISLs have the potential to ulteriorly enhance the SatCon capacity, however, acquisition, tracking, and pointing of the extremely narrow laser beams are still challenging in practice. Issues such as routing under time-varying network topology and creating multiple co-existing links among different orbital planes also require further research.}

\subsubsection{\bf Fast time-varing effect mitigation}
{Unlike terrestrial cellular systems, wireless channel in satellite communication is generally a Rician distributed LoS channel, and the satellite stations are in a state of high speed motion. Explicit distribution is conducive to channel modeling and algorithm design, but highly mobile satellites with fixed antenna inclination will introduce intractable issues, in which Doppler effect mitigation and synchronization are the most chanllenging tasks in broadband verticals.}

\subsubsection{\bf Access protocols for low latency}
{Now that the propagation latency is inevitable due to the round-trip delay between satellite and terrestrial nodes, novel mechanism should be developed to address the challenges of low-latency processing, particularly in massive IoT access networks. Grant-free random access (GFRA) allows active nodes to transmit their pilots and data to the BS without waiting for permissions\cite{grantfree}, which cuts down the overall latency, however, the corresponding spectrum resources consumption increases as the number of users increases, and the requisite accurate channel state information is difficult to obtain.}

\section{Conclusions}

In this article, we have given the development roadmap of LEO SatCons, and discussed the opportunities and key technologies of integrating LEO constellations with future cellular networks, where we have highlighted the vertical domains reshaped by LEO SatCons. In summary, as shown in Fig. \ref{cycle}, the key techniques involved in LEO SatCons are the pillars for achieving the superior KPIs and supporting various service requirements. {Superior} KPIs are the foundations for breeding and reshaping the various vertical applications, and revolutionary vertical applications will further bring profits for promoting the technical innovations of LEO constellations. It is highly anticipated that LEO SatCons will play a critical role in the reshaping of various vertical domains and further transform our society.
\section{Acknowledgement}
This work was supported by NSFC 62071044, Beijing Natural Science Foundation L182024.

\begin{IEEEbiographynophoto}
{Shicong Liu} received the B.S. degree in School of Information and Electronics in Beijing Institute of Technology, Beijing, China, in 2020, and is currently pursuing a M.S. degree in Beijing Institute of Technology, Beijing, China. 
\end{IEEEbiographynophoto}

\begin{IEEEbiographynophoto}
{Zhen Gao} (Member, IEEE) received the Ph.D. degree in communication and signal processing from the Tsinghua National Laboratory for Information Science and Technology, Department of Electronic Engineering, Tsinghua University, China, in July 2016. He is currently an Assistant Professor with the Beijing Institute of Technology.
\end{IEEEbiographynophoto}

\begin{IEEEbiographynophoto}
{Yongpeng Wu} (Senior Member, IEEE) received the Ph.D. degree in communication and signal processing from National Mobile Communications Research Laboratory, Southeast University, Nanjing, China, in 2013. He is currently a Tenure-Track Associate Professor with Shanghai Jiao Tong University, Shanghai, China.
\end{IEEEbiographynophoto}

\begin{IEEEbiographynophoto}
{Derrick Wing Kwan Ng} (Fellow, IEEE) received the Ph.D. degree from The University of British Columbia (UBC) in 2012. He is currently working as a Senior Lecturer and a Scientia Fellow at the University of New South Wales, Sydney, Australia.
\end{IEEEbiographynophoto}

\begin{IEEEbiographynophoto}
{Xiqi Gao} (Fellow, IEEE) received the Ph.D. degree in electrical engineering from Southeast University, Nanjing, China, in 1997. He joined the Department of Radio Engineering, Southeast University, in 1992, where he has been a Professor of information systems and communications, since 2001. 
\end{IEEEbiographynophoto}

\begin{IEEEbiographynophoto}
{Kai-Kit Wong} (Fellow, IEEE) received the Ph.D. degree in electrical and electronic engineering from The Hong Kong University of Science and Technology, Hong Kong, 2001. He is currently a Chair in Wireless Communications at the Department of Electronic and Electrical Engineering, University College London, U.K.
\end{IEEEbiographynophoto}

\begin{IEEEbiographynophoto}
	{Symeon Chatzinotas} (Senior Member, IEEE) received the Ph.D. degree in electronic engineering from the University of Surrey, Surrey, U.K., 2009. He is currently a Full Professor/Chief Scientist I and the Co-Head of the SIGCOM Research Group, SnT, University of Luxembourg.
\end{IEEEbiographynophoto}

\begin{IEEEbiographynophoto}
	{Bj\"{o}rn Ottersten} (Fellow, IEEE) received the Ph.D. degree in electrical engineering from Stanford University, Stanford, CA, USA, in 1989. He held research positions with Link\"{o}ping University, Stanford University, and the University of Luxembourg.
\end{IEEEbiographynophoto}
\clearpage
\begin{figure}[t]
	\centering
	\includegraphics[width=\textwidth]
	{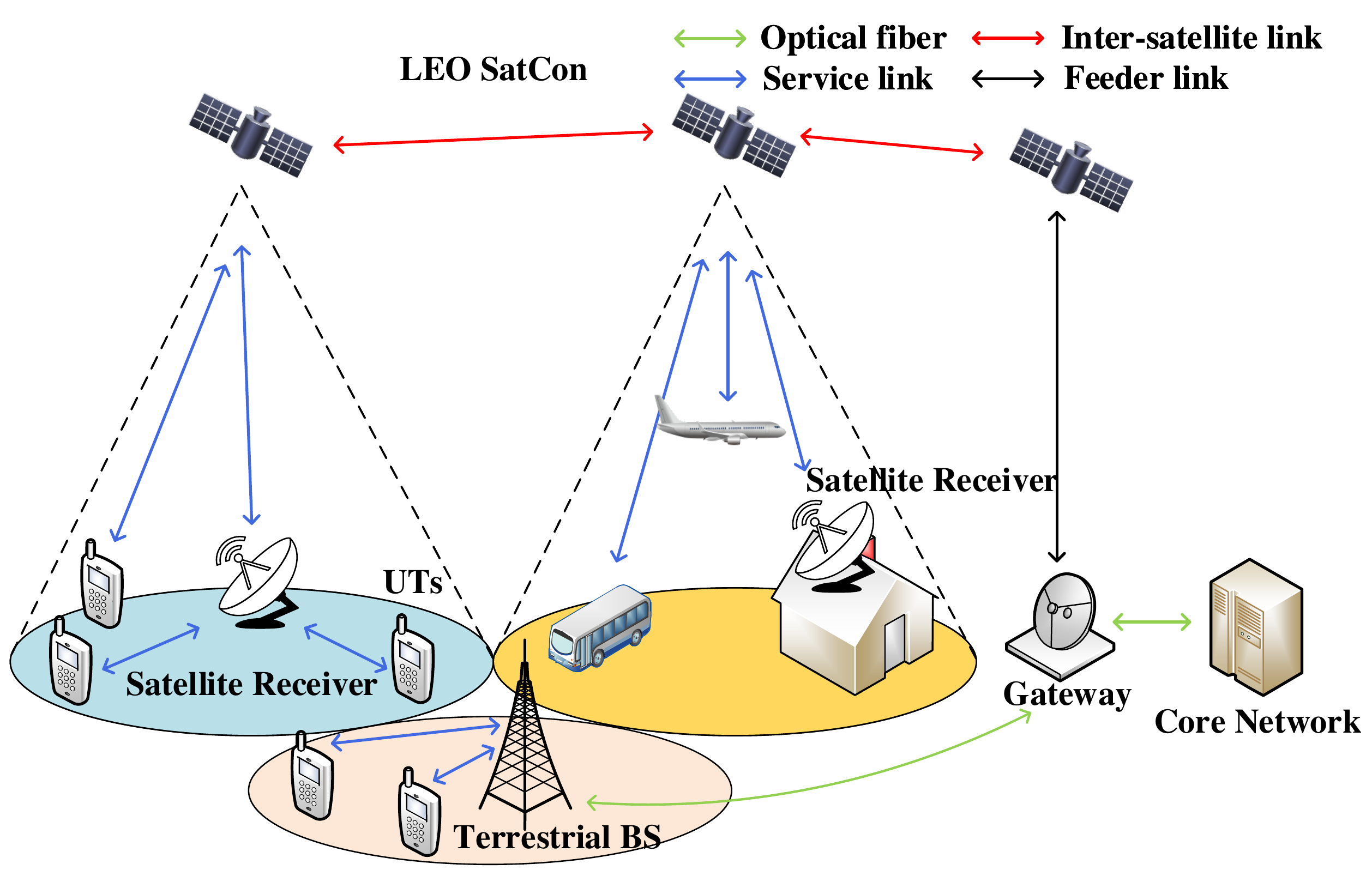}
	\caption{A schematic diagram of the network architecture of LEO SatCons and the common UT-SatCon access modes.}
\end{figure}
\clearpage
\renewcommand\arraystretch{1.25}
\begin{table*}[t]
	\centering
	\caption{Comparison of selected constellations}
	\label{comparison}
	\begin{threeparttable}
		\begin{tabular}{c|c|c|c|c|c|c}
			\hline
			{\bf Constellation}  & {\bf Regime} & {\bf Orbital height }& {\bf  Quantity }&{\bf  Bands } & {\bf Services}  &{\bf  Est. data rate}\\
			\hline
			Iridium Gen. 1  & LEO & $781$ km & 66 & L & Voice, data & 2.4 Kbps  \\
			\hline
			Globalstar  & LEO & $1414$ km& 48 & S, L  & Voice, data & $\sim$9.6 Kbps \\
			\hline
			Orbcomm Gen. 1  & LEO & $700\sim 800$ km & 36 & VHF & IoT \& M2M$^*$ communication & 2.4 Kbps\\
			\hline
			Skybridge  & LEO & $1457$ km & 64 & Ku  & Broadband Internet& 60 Mbps\\
			\hline
			Teledesic  & LEO & $1375$ km & 288(840) & Ka  & Broadband Internet& 64 Mbps \\
			\hline\hline
			Iridium NEXT  & LEO & $781$ km & 66 & L, Ka & Voice, data & 1.5 Mbps, 8 Mbps  \\
			\hline
			Orbcomm Gen. 2  & LEO & $700\sim 800$ km & 18 & VHF& IoT \& M2M communication & 4.8 Kbps\\
			\hline
			O3b  & MEO & $8063$ km & 20 & Ka& Broadband Internet & 500 Mbps\\
			\hline
			OneWeb  & LEO & $1200$ km & 648 & Ku & Broadband Internet & 400 Mbps\\
			\hline
			Starlink Gen. 1 & \multirow{2}{*}{LEO} & $335.9\sim 570$ & 11926 & V & \multirow{2}{*}{Broadband Internet} & \multirow{2}{*}{100 Mbps}\\
			Starlink Gen. 2 &  & {$328\sim 614$ km} km & 30000 &Ku, Ka, E &  & \\
			\hline
			Telesat Phase. 1  & \multirow{2}{*}{LEO} & \multirow{2}{*}{$1015\sim 1325$ km} & 298 & \multirow{2}{*}{Ku, Ka} & \multirow{2}{*}{Broadband Internet} & \multirow{2}{*}{-}\\
			Telesat Phase. 2  &   &   & 1373 & & & \\
			\hline
			Hongyan  & LEO & $1100$ km & 320 & L, Ka & Voice, broadband Internet & 100 Mbps\\
			\hline
			Kuiper  & LEO & $590\sim 630$ km & 3236 & Ka & Broadband Internet & -\\
			\hline
		\end{tabular}
		\begin{tablenotes}
			\footnotesize
			\item[*$\ \ $]: Internet of Things and machine to machine.
		\end{tablenotes}
	\end{threeparttable}
\end{table*}
\clearpage
\begin{figure*}[t]
	\centering
	\includegraphics[width=\textwidth]
	{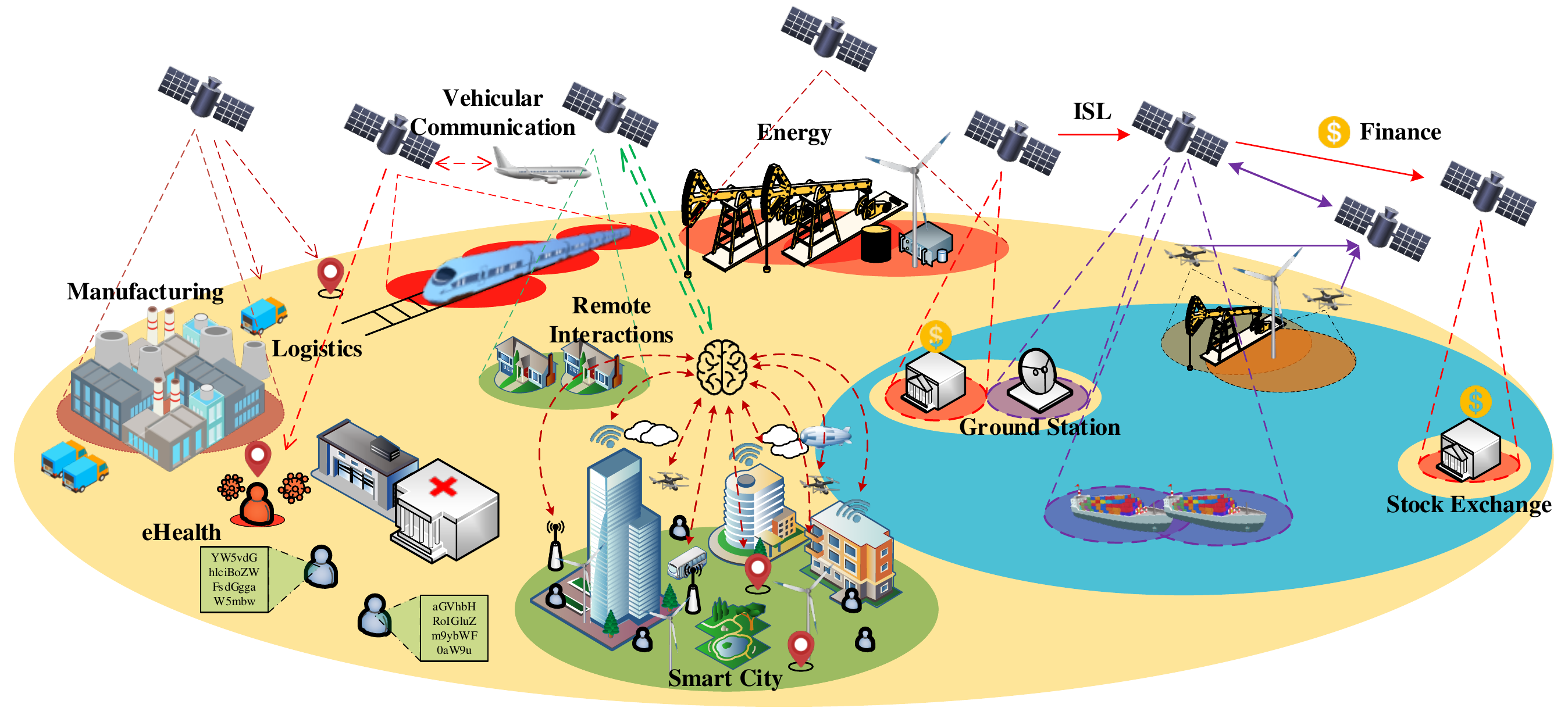}
	\caption{Ambitious vision of various vertical domains reshaped by LEO SatCons.}
\end{figure*}
\clearpage
\begin{figure}[t]
	\centering
	\includegraphics[width=\textwidth]
	{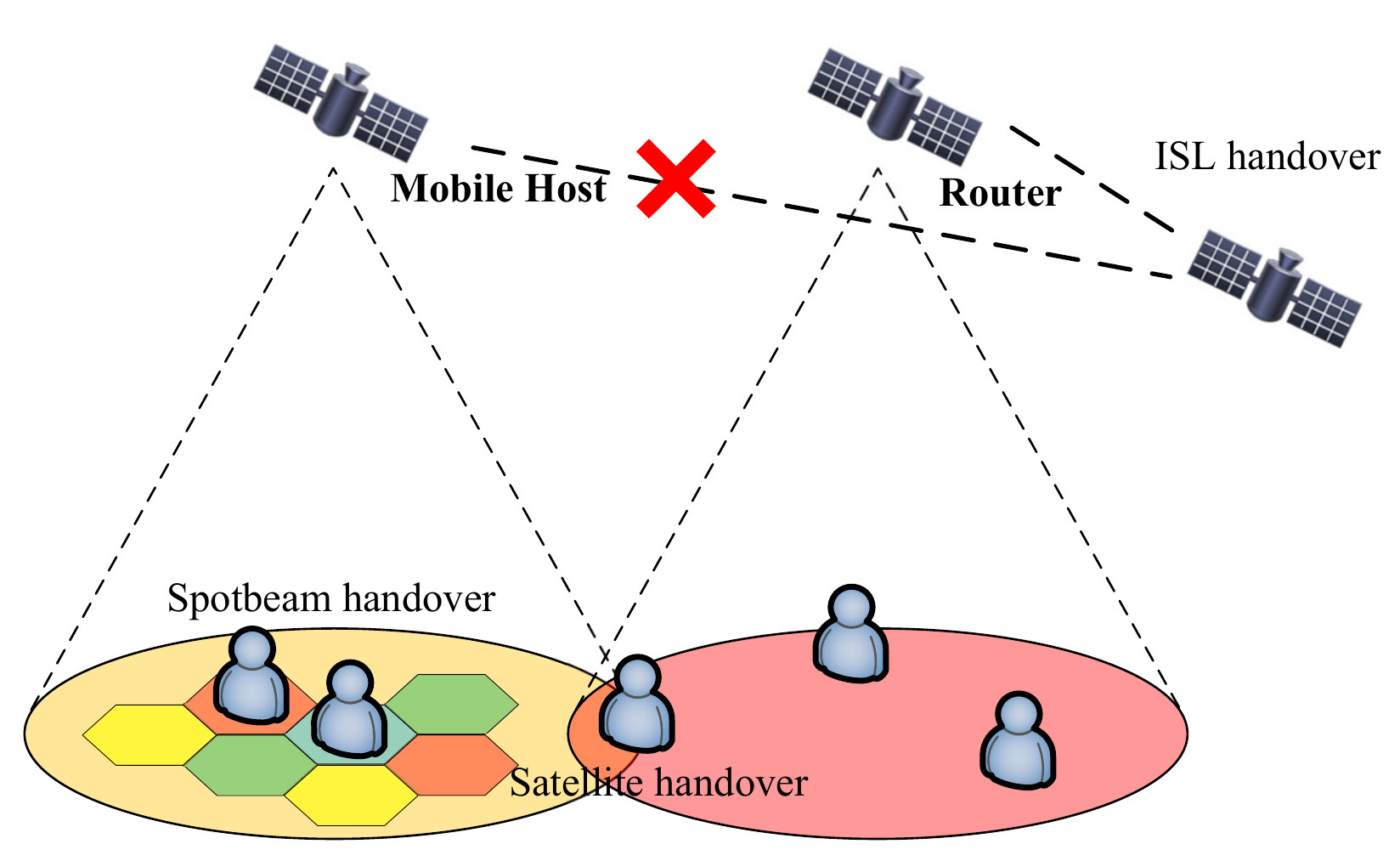}
	\caption{Typical link layer handover schemes.}
\end{figure}
\clearpage
\begin{figure*}[t]
	\centering
	\includegraphics[width=\textwidth]
	{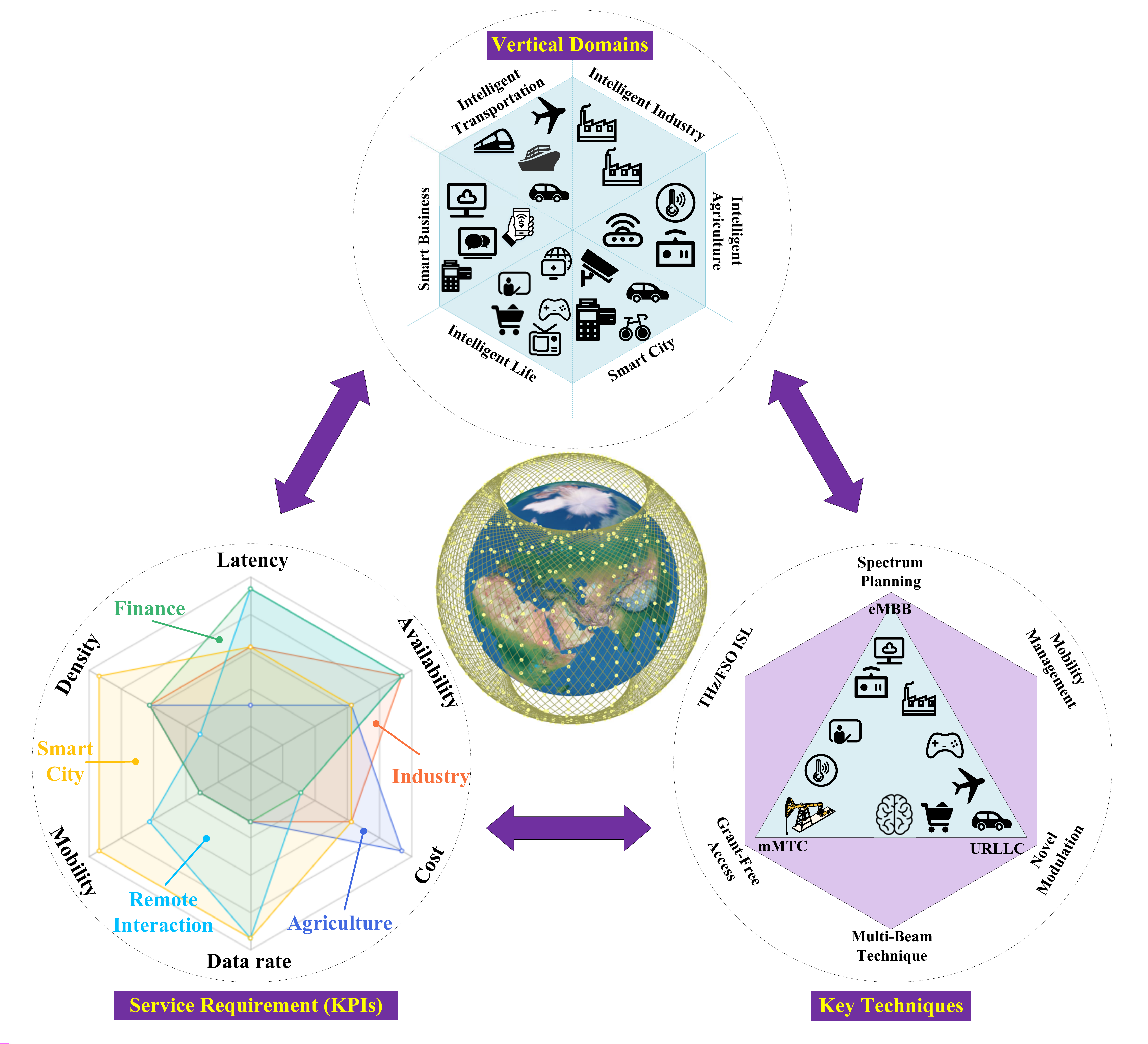}
	\caption{Relationship between vertical domains, service requirements, and key techniques.}
\end{figure*}
\end{document}